\documentclass[12pt,onecolumn,floatfix,amsmath,nofootinbib,amssymb,floatfix,aps,prd,showpacs]{revtex4}
\usepackage{slashed}
\usepackage[level]{datetime}
\usepackage{graphicx,color,dcolumn,booktabs,bm}
\usepackage{longtable,lscape}
\usepackage{txfonts}
\usepackage{subfigure}
\usepackage{float}
\usepackage{makecell}
\usepackage{tabularx}
\usepackage{array}
\usepackage{verbatim}
\usepackage{threeparttable}
\usepackage{ragged2e}
\newcolumntype{P}[1]{>{\centering\hspace{0pt}}p{#1}}
\newcolumntype{Z}{>{\centering\let\newline\\\arraybackslash\hspace{0pt}}x}

\usepackage{booktabs}
\usepackage{overpic}
\usepackage{amssymb}
\usepackage{indentfirst}
\usepackage{feynmf}   
\usepackage{slashed}  
\usepackage{cases}
\usepackage{color}
\usepackage{multirow}
\usepackage{epstopdf}
\usepackage{longtable}
\usepackage{indentfirst}
\usepackage{graphicx,color,dcolumn,booktabs,bm}
\usepackage[colorlinks,
            citecolor=blue,
            anchorcolor=red,
            menucolor=red,
            linkcolor=red,
            filecolor=red,
            runcolor=red,
            urlcolor=blue,
            frenchlinks=red]{hyperref}
\graphicspath{{Figures/}} %
\allowdisplaybreaks
\usepackage{CJK}
\usepackage{epstopdf}
\begin{document}

\title{$D\rightarrow \pi$ transitions from QCD Light-Cone Sum Rules with the chiral currents}
\author{Di Gao$^{1,2,3}$}
\email{digao@impcas.ac.cn}
\author{Jiangshan Lan$^{2,5,6}$}
\email{jiangshanlan@impcas.ac.cn}
\author{Duojie Jia$^{4,7}$ \thanks{}}
\email{jiadj@nwnu.edu.cn; Corresponding author}
\author{Xingbo Zhao$^{2,5,6}$}
\email{xbzhao@impcas.ac.cn}
\author{Yanjun Sun$^{1,7}$}
\email{sunyanjun@nwnu.edu.cn}
\affiliation{$^1$Institute of Theoretical Physics, College of Physics and Electronic Engineering, Northwest Normal University, Lanzhou 730070, China \\
$^2$Institute of Modern Physics, Chinese Academy of Sciences, Lanzhou 730000, China  \\
$^3$Lebedev Physical Institute, Russian Academy of Sciences, Leninsky Prospekt 53, 119991, Moscow, Russia  \\
$^4$General Education Center, Qinghai Institute of Technology, 810000, Xining, China \\
$^5$School of Nuclear Science and Technology, University of Chinese Academy of Sciences, Beijing 100049, China \\
$^6$CAS Key Laboratory of High Precision Nuclear Spectroscopy, Institute of Modern Physics, Chinese Academy of Sciences,  Lanzhou 730000, China\\
$^7$Lanzhou Center for Theoretical Physics, Lanzhou University, Lanzhou 730070, China  \\
}

\begin{abstract}
We present a systematic study of the transition form factors for the semileptonic decays $D\rightarrow Pe^{+}v_{e}$ and $D\rightarrow P\mu^{+}\bar{v}_{\mu}$ using QCD light-cone sum rules with chiral currents, with emphasis on the nonperturbative structure of the pion.  The distribution amplitudes of the meson $\pi$, including twist-2 and higher-twist components, are analyzed and incorporated in our computation, employing both Gegenbauer polynomial expansions and the Basis Light-Front Quantization (BLFQ) method, where the latter yields a distribution that rapidly converges to the asymptotic form. Two relations linking transition form factors are derived and found to be consistent with the chiral symmetry of QCD.  Our numerical predictions for the $D$-to-pion form factors agee well with BES\uppercase\expandafter{\romannumeral3} measurements and lattice QCD calculations. Furthermore, the differential decay widths obtained thereby for the $D$-to-pion decay also show consistency with the BES\uppercase\expandafter{\romannumeral3} data in the high-momentum-transfer ($q$) region with the $q^{2}>1\ \rm{GeV^2}$, implying new opportunities for the precise extraction of the CKM matrix element and the exploration of physics beyond the Standard Model.
\end{abstract}
\maketitle
\section{Introduction}
Heavy to light decays, encompassing both inclusive \cite{Fael:2019umf,Calderon:2007nw,Bauer:2004ve,Belle:2004afj} and exclusive \cite{Nayak:2022qaq,Ivanov:2019nqd,Biancofiore:2014uba,Barik:2009zzb} processes, provide substantial insights into the dynamics of weak and strong interactions, and play a unique role in extracting the Cabibbo-Kabayashi-Maskawa (CKM) matrix elements (e.g., $|V_{cd}|$) and probing phenomena beyond the standard model (SM). Generally, the inclusive decays of heavy mesons are more challenging experimentally but better understood theoretically  than exclusive decays. In contrast, exclusive decays offer higher experimental precision but pose significant theoretical challenges for nonperturbative methods to compute form factors (FFs) that encode long-distance dynamics.
\par
On the one hand, recent high-statistics measurements of branching fractions and form factors for $D/B$-meson decays to pseudoscalar mesons, such as those from the Beijing Spectrometer (BES\uppercase\expandafter{\romannumeral3}) \cite{BESIII:2017ylw,BESIII:2017ikf,BESIII:2018nzb,BESIII:2018eom,BESIII:2019gsm,BESIII:2019qci,BESIII:2018xre,BESIII:2018jjm,BESIII:2018qmf}, have significantly improved precision. On the other hand, the heavy-to-light decays, including leptonic, semileptonic and non-leptonic modes, have been analyzed by using lattice Quantum Chromodynamics (LQCD) \cite{Abada:1993dh,APE:1994kxx,Gusken:1995ni,Nieves:1994cq,Flynn:1997ca,DelDebbio:1997ite,Abada:1999xd,Abada:2000ty,FermilabLattice:2004ncd,Bernard:2009ke,Al-Haydari:2009kal,DiVita:2010mlb,Na:2011mc,FermilabLattice:2019ycs}, quark model \cite{Wirbel:1985ji,Isgur:1988gb,Sun:1990fa,Scora:1995ty,Beyer:1998ka,Melikhov:2000yu}, QCD sum rule \cite{Belyaev:1993wp,Khodjamirian:1997lay,Ball:1998kk,Khodjamirian:2000ds}, heavy quark symmetry (HQS) and heavy quark effective field theory (HQEFT) \cite{Wu:1992zw,Wang:1999zd,Wang:2000sc}. These 
theoretical efforts have significantly promoted the calculation of the corresponding form factors.
\par
Purpose of this work is to explore the form factors (FFs) for the semileptonic decays $D\rightarrow P$ using the light-cone QCD sum rule (LCSR), where the relevant distribution amplitudes (DAs) are obtained from approach of the basis light-front quantization (BLFQ). Our computed DAs are compared to other theoretical predictions. Next, we analyze the semileptonic decay process and reveal simple relations among the FFs for the $D \rightarrow\pi$ transition. Finally, we present the differential decay width and make a direct comparison of them with the experimental data reported by  BES\uppercase\expandafter{\romannumeral3} \cite{BESIII:2017ylw}.
\par
As illustrated in Fig. \ref{phi1-4}, many studies adopt a Gegenbauer expansion for the DA truncated at a finite order (typically $a_2$ or $a_4$). This approach naturally prompts two critical questions: (a) whether the expansion converges at the chosen truncation order, and (b) whether the extracted coefficients are biased by method-specific parameters (e.g., the Borel parameter and continuum threshold in QCD sum rules). To address these questions, we employ the BLFQ method as a model-independent benchmark. By reconstructing the Gegenbauer coefficients from the BLFQ-derived DA, we observe excellent consistency between the reconstructed and reference DAs, confirming the robustness of the Gegenbauer expansion. This indicates that the variations among previous predictions stem primarily from parameter choices rather than the expansion framework itself. We discuss this point further in Sec.~II.
\par
The paper is organized as follows: Sect. \uppercase\expandafter{\romannumeral2} presents a comparison of the light-cone distribution amplitudes (LCDAs) for the pseudoscalar meson $\pi$ based on various theoretical predictions. In Sect. \uppercase\expandafter{\romannumeral3}, we derive the form factors and differential branching fractions for the $D\rightarrow Pl\bar{\nu}_{l}$ semileptonic decays using the LCSR method. Sect. \uppercase\expandafter{\romannumeral4} is devoted to the numerical analysis, where we compare our results with other theoretical predictions and experimental data. The paper concludes with a summary in Sect. \uppercase\expandafter{\romannumeral5}.

\section{The distribution amplitude of Pion}
Understanding the intrinsic structure of pions is crucial in exploring the low-energy QCD. Composed of valence up and down quarks and antiquarks, pions are Nature’s one of most fundamental particle known as pseudo-Nambu-Goldstone bosons. They are intrinsically linked to chiral symmetry breaking and QCD bound states in strong interactions. Furthermore, since pions possess lepton-like masses, studying them may shed light on the nonperturbative mechanisms behind the generation of hadron masses. Therefore, elucidating the pion distribution amplitude (DA), which describes the probability amplitude of finding the pion in a quark-antiquark Fock state on the light cone, is of great significance in subatomic physics. In this section, we introduce the DAs calculated via both conventional and innovative approaches. Thereby, we compare the DA directly obtained from the BLFQ method with the DA expressed in terms of Gegenbauer polynomials (derived from the BLFQ method by assuming it represents the true physical DA and using the definition of Gegenbauer moments) to verify the convergence of the Gegenbauer expansion for the pion.

\subsection{The DAs as a sum of Gegenbauer polynomial\label{secA}}
The light-cone distribution amplitudes of the pseudoscalar meson $\pi$ made up of one light quark $q_{1}$ and light antiquark $\bar{q}_{2}$ 
, have the form of \cite{Radyushkin:1977gp,Braun:1989iv,Huang:2004tp}
\begin{equation}
\begin{aligned}
\langle P(p)|\bar{q}_{2}(x)\gamma_{\mu}\gamma_{5}q_{1}(0)|0\rangle&=-ip_{\mu}f_{P}\int_{0}^{1}du e^{iup\cdot x}\phi_{P}(u),\\
\langle P(p)|\bar{q}_{2}(x)i\gamma_{5}q_{1}(0)|0\rangle&=f_{P}m_{P}\int_{0}^{1}du e^{iup\cdot x}\phi_{P}^{S}(u),\\
\langle P(p)|\bar{q}_{2}(x)\sigma_{\mu\nu}\gamma_{5}q_{1}(0)|0\rangle&=-i(p_{\mu}x_{\nu}-p_{\nu}x_{\mu})f_{P}m_{P}\int_{0}^{1}du e^{iup\cdot x}\phi_{P}^{\sigma}(u),\\
\end{aligned}
\end{equation}
where $\phi_{P}$ is the twist-2 light-cone DA, $\phi_{P}^{S}$ and $\phi_{P}^{\sigma}$ are the twist-3 light-cone DAs, $u$($\bar{u}=1-u$) denotes the momentum fraction carried by $\bar{q}_{2}$($q_{1}$) quark in the pseudoscalar meson, and $f_{P}$ is the decay constant of pseudoscalar meson, which is defined by
\begin{equation}
\langle P(p)|\bar{q}_{2}\gamma_{\mu}\gamma_{5}q_{1}|0\rangle=-if_{P}p_{\mu}.\notag
\end{equation}

We discuss only the twist-2 light-cone DA $\phi_{P}(u)$ as the twist-3 light-cone DAs $\phi_{P}^{S}(u)$ and $\phi_{P}^{\sigma}(u)$ vanish when we use chiral currents. In addition, we do not take the higher twist distribution amplitudes into account. The corresponding renormalization group (RG) equation at one loop \cite{Lepage:1980fj,Efremov:1979qk,Lepage:1979zb,Mueller:1994cn,Belyaev:1994zk} implies the series expansion of the DA of pion in terms of Gegenbauer polynomials, 
\begin{equation}
\phi(u)=6u(1-u)\sum_{n}a_{n}C_{n}^{\frac{3}{2}}(2u-1),
\label{defphi}
\end{equation}
based on the approximate conformal symmetry of QCD with light quarks. Here, the Gegenbauer coefficients $a_{n}$ determine the deviation of $\phi_{P}(u)$ from its asymptotic form $6u(1-u)$, and $C_{n}^{\frac{3}{2}}$, known as Gegenbauer polynomials, can be normalized by
\begin{equation}
\int du C_{n}^{\frac{3}{2}}(2u-1)C_{m}^{\frac{3}{2}}(2u-1)=\frac{(n+2)(n+1)}{4(2n+3)}\delta_{mn}.
\end{equation}

In addition, Gegenbauer coefficients $a_{n}$ and meson decay constant $f_{P}$ run as the energy scale $\mu$ changes, whose evolution with respect to $\mu$ is given by the renormalization group equations \cite{Braun:2003rp,Cheng:2005nb}
\begin{equation}
\begin{aligned}
a_{n}(\mu)&=a_{n}(\mu_{0})(\frac{\alpha(\mu_{0})}{\alpha(\mu)})^{-(\gamma_{(n)}+4)/b},\notag\\
f_{P}(\mu)&=f_{P}(\mu_{0})(\frac{\alpha(\mu_{0})}{\alpha(\mu)})^{4/b},
\end{aligned}
\end{equation}
where the one-loop anomalous dimensions are given by \cite{Gross:1974cs,Shifman:1980dk}
\begin{equation}
\gamma_{(n)}=C_{F}[1-\frac{2}{(n+1)(n+2)}+4(\sum_{j=2}^{n+1}\frac{1}{j})],\notag
\end{equation}
with
\begin{equation}
C_{F}=\frac{N_{c}^{2}-1}{2N_{c}}.\notag
\end{equation}

Given these relations, we are able to explore the Gegenbauer coefficients $a_{n}$ of the pion mesons, which is closely related to moments of pion at the energy scale $\mu=1\ \rm{GeV}$. The obtained results are shown in Table \ref{T1}.  

\begin{figure}[H]
\centering
\includegraphics[height=6.5cm,width=8.5cm]{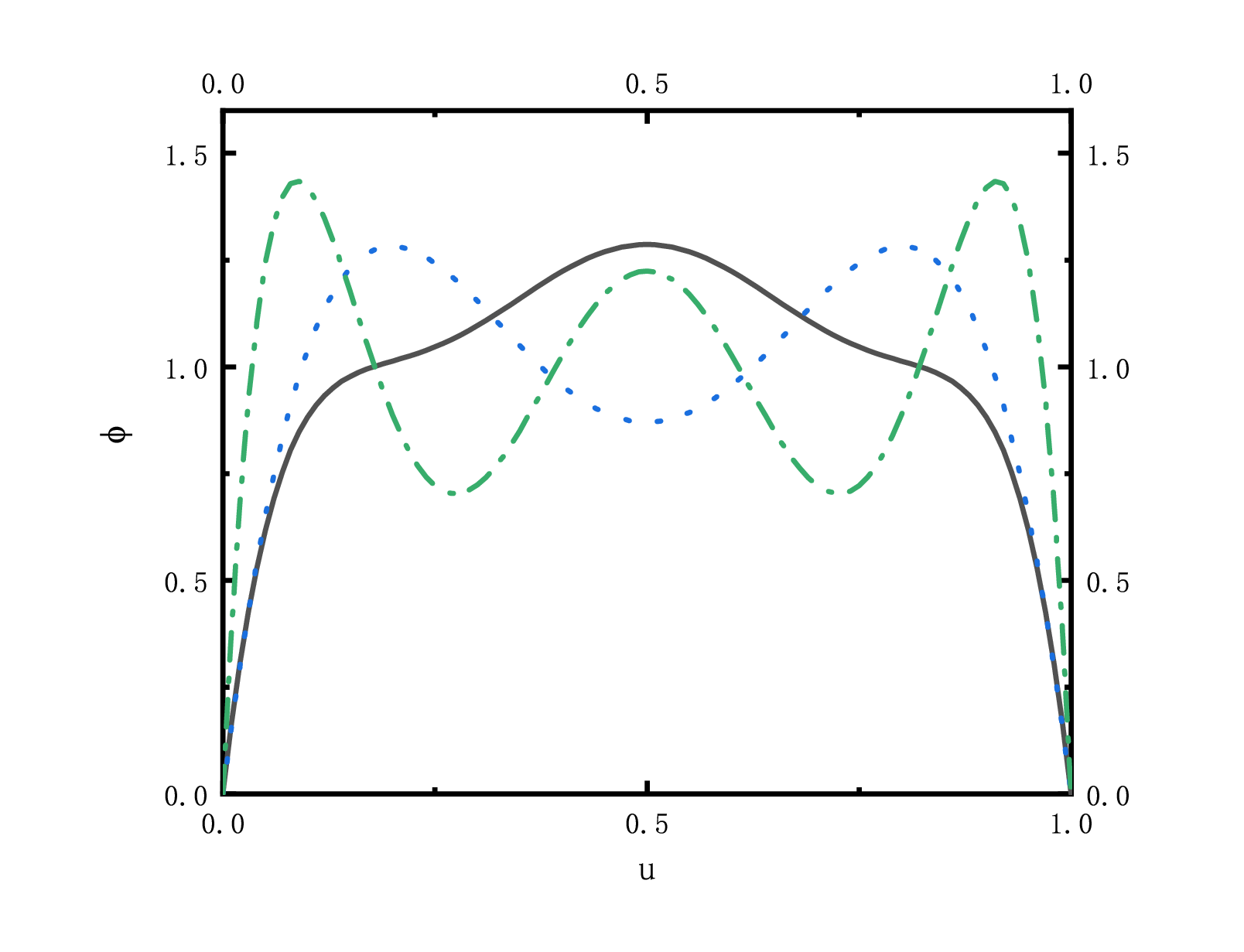}
\caption{The distribution amplitudes via Gegenbauer expansions for pion. The solid, dot and dash dot lines correspond to $\{a_{2,4}\}=\{0.17,0.06\}$ \cite{Khodjamirian:2011ub},  $a_{2}=0.28$ \cite{Ball:2006yd} and $\{a_{2,4}\}=\{0.41,0.23\}$ \cite{Belyaev:1994zk,Khodjamirian:1997lay,Khodjamirian:2000ds,Braun:1988qv}, respectively}
\label{phi1-4}
\end{figure}

\begin{table}[H]
\caption{Comparison of Gegenbauer coefficients $a_2$ and $a_4$ of DA predicted for pion mesons. }
  \centering
   \begin{threeparttable}
    \label{T1}
	 \begin{tabular}{p{6cm}<{\centering}p{5cm}<{\centering}p{5cm}<{\centering}}
      \hline
      \hline
        ~                          &  $a_{2}$ & $a_{4}$ \\
       \hline
       Khodjamirian \cite{Khodjamirian:2011ub} & 0.17     & 0.06  \\
       Duplancic \cite{Duplancic:2008ix}    & 0.16     & 0.04  \\
       Ball \cite{Ball:2006yd}         & 0.28     & ~  \\
       Belyaev, Braun et al \cite{Belyaev:1994zk,Khodjamirian:1997lay,Khodjamirian:2000ds,Braun:1988qv}      & 0.41     & 0.23 \\
     \hline
     \hline
   \end{tabular}
   \end{threeparttable}
  \centering
\end{table}
Here, the distribution amplitudes in Ref. \cite{Khodjamirian:2011ub} are derived at next-to-leading order (NLO) via QCD factorization, whereas the Gegenbauer coefficients in Refs. \cite{Duplancic:2008ix,Ball:2006yd,Belyaev:1994zk,Khodjamirian:1997lay,Khodjamirian:2000ds,Braun:1988qv,Wu:2006rd} are obtained from QCD sum rules. 
Given these  Gegenbauer coefficients and polynomials, one can apply Eq. (\ref{defphi}) to compute the distribution amplitudes. The obtained results is plotted in FIG.~\ref{phi1-4}

\subsection{The DAs with BLFQ method\label{secB}}

In the light-front form of  QCD field theory, the hadrons can be described by the light-front wave functions  (LFWFs), which are given, for the case of mesons, by \cite{Lan:2021wok}
\begin{equation}
\begin{aligned}
\Psi^{N,M_{J}}_{\{x_{i};\overrightarrow{p}_{\bot};\lambda_{i}\}}=\sum_{n_{i};m_{i}}\psi^{N}(\{\overline{\alpha_{i}}\})\prod_{i=1}^{N}\phi_{n_{i};m_{i}}(\overrightarrow{p}_{\bot};b),
\end{aligned}
\end{equation}
where $\psi^{N=2}(\{\overline{\alpha_{i}}\})$ and $\psi^{N=3}(\{\overline{\alpha_{i}}\})$ are the components of the eigenvectors relevant to the Fock sectors of quark-antiquark configuration $|q\bar{q}\rangle$ and  the hybrid meson configuration  $|q\bar{q}g\rangle$, with $g$ the valence gluon, $\{\overline{\alpha}\}=\{x, n, m, \lambda\}$, $x$, $n$, $m$, $\lambda$ are the longitudinal momentum fraction, the radial, angular and the additional quantum number,  respectively, and $\phi_{n_{i};m_{i}}(\overrightarrow{p}_{\bot};b)$ is the two dimensional harmonic oscillator basis functions which can be obtained by diagonalizing the full Hamiltonian matrix of QCD, e.g.,  in the
BLFQ \cite{Vary:2009gt,Xu:2021wwj,Zhao:2014xaa,Wiecki:2014ola}.
\par
We use the truncation $N_{max} =14$, and the six model parameters, all of which are summarized in Table \ref{T2}. With these fixed inputs of the parameters, one can generate the mass spectra of light mesons \cite{Lan:2021wok}. Here, the model parameters include the quark mass $m_{q}$, the effective mass of gluon $m_{g}$, the harmonic oscillator
scale parameter $b$, the strength of the confinement $\kappa$, the two parameters $m_{f}$ and $g$ (independent of quark mass) in the vertex interaction and coupling constant.

\begin{table}[H]
    \centering
    \caption{The model parameters for the truncation $N_{max}=14$ \cite{Lan:2021wok}. All are in units of $\rm{GeV}$ except $g$, which is dimensionless.}
	\begin{tabular}{p{2cm}<{\centering}p{2cm}<{\centering}p{2cm}<{\centering}p{2cm}<{\centering}p{2cm}<{\centering}p{2cm}<{\centering}}
		\hline
        \hline
        $m_{q}$ &  $m_{g}$&  $b$  & $\kappa$ & $m_{f}$ & $g$ \\
        \hline
        0.39 & 0.60 & 0.29 & 0.65 & 5.69 & 1.92\\
        \hline
        \hline
        \label{T2}
	\end{tabular}
\end{table}
\par
The DAs are defined in  terms of the light-like separated gauge invariant vacuum-to-meson matrix elements. In the light-front formalism, the leading-twist DAs $\phi_{P}(u, \mu)$ in the light-cone gauge for a pseudoscalar are defined by \cite{Lepage:1980fj}
\begin{equation}
\begin{aligned}
\langle0|\bar{q}_{2}(z)\gamma^{+}\gamma_{5}&q_{1}(-z)|P(p)\rangle=if_{P}p^{+}\int_{0}^{1}dxe^{ip^{+}z^{-}(u-\frac{1}{2})}\phi_{P}(u).
\label{BLFQp}
\end{aligned}
\end{equation}
The DAs of pseudoscalar states are linked to the LFWF $\Psi$  in the form of  \cite{Li:2015zda,Tang:2018myz,Jia:2018ary,Mondal:2019jdg,Lan:2019rba,Lan:2019vui,Qian:2020utg,Li:2021jqb,Lan:2021wok,Lan:2022blr}
\begin{equation}
\phi_{P}(u)=\bar{f}_{P}\frac{2\sqrt{2N_{C}}}{N_{F}}\int[d^{2}p_{\bot}]\Psi^{\uparrow\downarrow+\downarrow\uparrow}(u,p_{\bot}),
\label{BLFQDA}
\end{equation}
with
\begin{equation}
N_{F}=\sum_{s_{1}s_{2}}\int[d^{3}p_{1}][d^{3}p_{2}]\Psi^{s_{1}s_{2}*}(p_{1},p_{2})\Psi^{s_{1}s_{2}}(p_{1},p_{2}),\notag
\end{equation}
where, $N_{C}$ is the number of colors, $p_{1,2}$ are moments of quarks $q_{1,2}$, with spin $s_{1,2}$, $s_{1}$ and $s_{2}$ are spins. 
In the Nambu-Jona-Lasinio model with Hamiltonian of the Basis Light front Quantization form, which we refereed as the BLFQ-NJL model, we use $N_{C} = 3$ in this work. Recall that in the BLFQ-NJL model the flavor wave function of the neutral Pion is the only difference compared to the wave function of the charged Pion. Therefore, one can compute the distribution amplitude of Pion, with the obtained results shown in FIG. ~\ref{phicd}. In the domain $m^{2}_{\pi}/ Q^{2}\backsimeq0$ \cite{Lepage:1979zb,Efremov:1979qk,Lepage:1980fj}, the DA of the pion is known to be $\phi(u)=6u(1-u)$, and this prediction is reproduced by the obtained result of BLFQ basically (FIG. ~\ref{phicd}).

\subsection{The DAs as a sum of Gegenbauer polynomial with BLFQ method}
\begin{figure}[H]
\centering
\includegraphics[height=6.5cm,width=8.5cm]{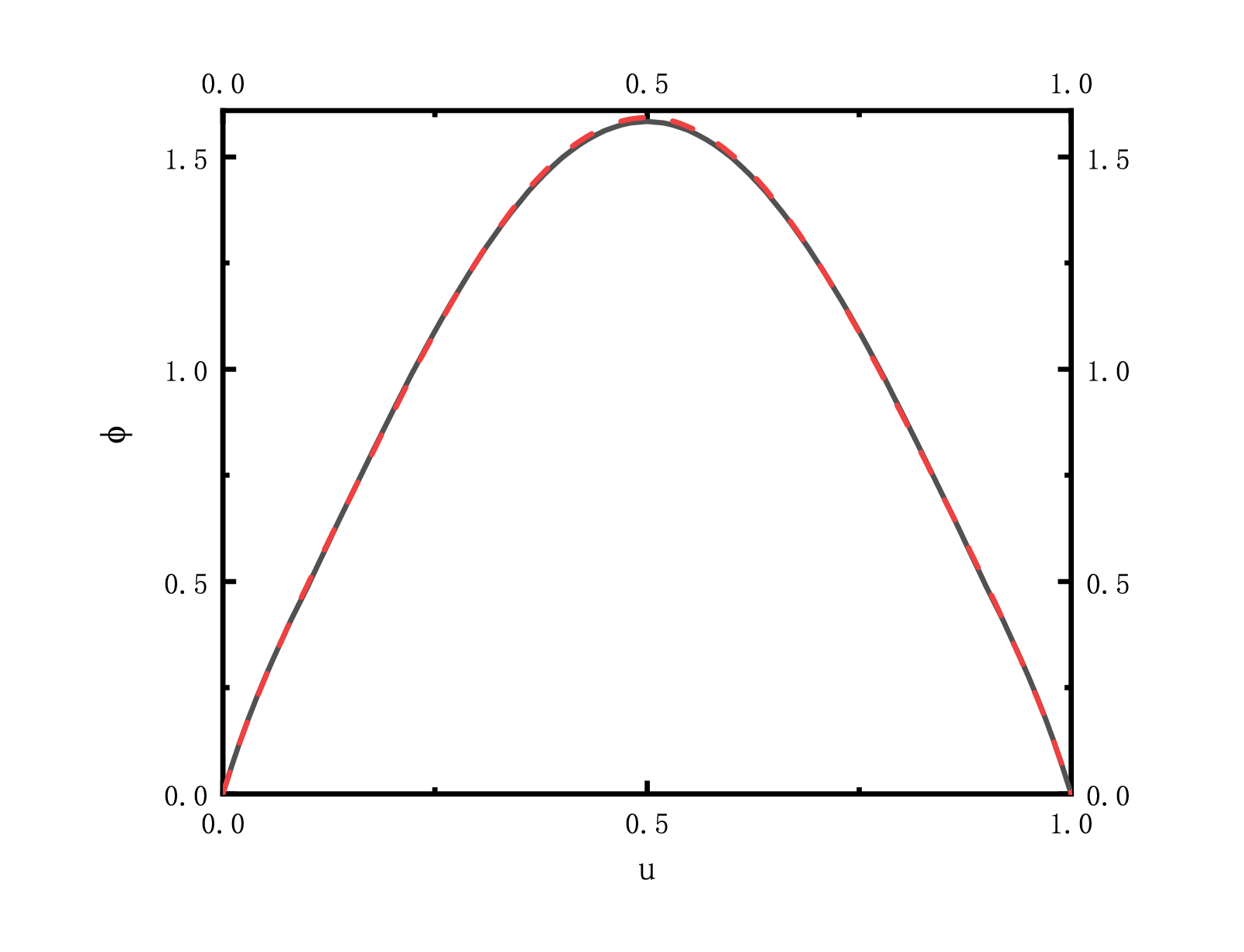}
\caption{The two pion DAs based on and with the BLFQ method (solid and dashed lines, respectively). They are almost coincident.}
\label{phicd}
\end{figure}

Given that the precise determination of the DAs of the pion, based on the method of the BLFQ, are precisely determine, the relationship between the moments and the DAs can be established without recourse to any approximations, as follows:
\begin{equation}
\langle\xi^{n}\rangle=\int_{0}^{1}(2x-1)^{n}\phi_{P}(x),
\end{equation}
where $\phi_{P}(x)$ is extracted by Eq. (\ref{BLFQDA}). Subsequently, one can evaluate the Gegenbauer coefficients at the scale $\mu = 1\ \rm{GeV}$, and  obtains the following sequence of the Gegenbauer coefficients 
\begin{equation}
\{a_{0}, a_{2}, a_{4}, a_{6}, a_{8}, a_{10}, \cdots\}=\{1, -0.0283, 0.0130, 0.0024, 0.0012, -0.0010, \cdots\}.\notag
\end{equation}
It is evident that the Gegenbauer coefficients show a clear trend of convergence as the order n increases. The distribution amplitude obtained from the BLFQ method is shown in FIG. \ref{phicd}.
In addition, we perform a comparative analysis of the DA based on and that with the BLFQ method. One can juxtapose the DAs obtained with two approaches and present the comparison in FIG. \ref{phicd}. 

The exact overlap of the two lines representing the DAs based on and from(with) the BLFQ method suggests a high degree of consistency between the distribution amplitudes obtained by the two methods.

\section{Form factors}
To calculate the form factor of the decay $D\rightarrow P$, we define the two-point correlation function
\begin{equation}
\begin{aligned}
\Pi_{1\mu}(p,q)&=i\int d^{4}xe^{iqx}\langle P(p)|T\{\bar{q}_{2}(x)\gamma_{\mu}(1+\gamma_{5})c(x),\bar{c}(0)i(1+\gamma_{5})q_{1}(0)\}|0\rangle,\\
\Pi_{2\mu}(p,q)&=i\int d^{4}xe^{iqx}\langle P(p)|T\{\bar{q}_{2}(x)\sigma_{\mu\nu}(1+\gamma_{5})q^{\nu}c(x),\bar{c}(0)i(1-\gamma_{5})q_{1}(0)\}|0\rangle,
\end{aligned}
\label{CF1}
\end{equation}
where $P$ is light pseudoscalar mesons. The momenta $q$ is defined by $q=p_{D}-p$, where $p_{D}$ and $p$ are the four-momentum of the initial and final meson states, respectively. Here, the correlation function $\Pi_{1\mu}$ is associated with the vector 
current contribution, while $\Pi_{2\mu}$ accounts for the tensor current effects.
\par
To derive the form factors for the $D\rightarrow P$ semileptonic decays, we shall use the LCSR method to deal with the correlation function on two sides.  The standard procedure for this are 
\begin{enumerate}
\item On the phenomenological side, one inserts a set of hadron complete setd of initial meson states in the middle of two currents in the correlation function.
\item On the theoretical side, one uses operator product expansion (OPE) upon correlation functions.
\end{enumerate}
Then, equating the two representations obtained via the above procedure and applying the Borel transformation to suppress the contributions from higher states and the continuum, one can obtain the sum rule expressions for the form factors. The following subsections are devoted to the details of the relevant analysis. 
\subsection{The phenomenological side\label{sec3A}}
For the correlation functions in Eq. (\ref{CF1}), one can insert a set of complete sets of mesons, and then isolate the mass term of the pseudoscalar $D$ mesons. The results are obtained as, upon applying Fourier transformation to them,
\begin{equation}
\begin{aligned}
\Pi_{1\mu}(p,q)&=\frac{\langle P(p)|\bar{q}_{2}\gamma_{\mu}c|D\rangle \langle D|\bar{c}i\gamma_{5}q_{1}|0\rangle}{m^{2}_{D}-(p+q)^{2}}+higher \quad states,\\
\Pi_{2\mu}(p,q)&=\frac{\langle P(p)|\bar{q}_{2}\sigma_{\mu\nu}(1+\gamma_{5})q^{\nu}c|D\rangle \langle D|\bar{c}i\gamma_{5}q_{1}|0\rangle}{m^{2}_{D}-(p+q)^{2}}+higher \quad states.
\label{CF}
\end{aligned}
\end{equation}
where the hadronic matrix elements $\langle P(p)|\bar{q}_{2}\gamma_{\mu}c|D\rangle$ and $\langle P(p)|\bar{q}_{2}\sigma_{\mu\nu}(1+\gamma_{5})q^{\nu}c|D\rangle$ can be parameterized by a set of $q$-dependent functions in the following forms: \cite{Khodjamirian:1997ub,Ball:2004ye,Duplancic:2008ix,Khodjamirian:2009ys}:
\begin{equation}
\begin{aligned}
\langle P(p)|\bar{q}_{2}\gamma_{\mu}c|D\rangle&=2f_{+}^{P}(q^{2})p_{\mu}+[f_{+}^{P}(q^{2})+f_{-}^{P}(q^{2})]q_{\mu},\\
\langle P(p)|\bar{q}_{2}\sigma_{\mu\nu}(1+\gamma_{5})q^{\nu}c|D\rangle&=[2p_{\mu}q^{2}-2q_{\mu}(p\cdot q)]\frac{if_{T}^{P}(q^{2})}{m_{D}+m_{P}},
\label{FFp}
\end{aligned}
\end{equation}
where $f_{+}^{P}$ and $f_{-}^{P}$ are the transition form factors of the semileptonic decays $D\rightarrow P$, $f_{T}^{P}(q^{2})$ is the penguin form factor, $m_{D}$ and $m_{P}$ are the mass of $D$ and pseudoscalar mesons, $m_{c}$ and $m_{q_{1}}$ are the mass of charm quark ($c$) and light quark, respectively.
The $D$ meson decay constant $f_{D}$ is defined by the other matrix element:
\begin{align}
\langle D|\bar{c}i\gamma_{5}q_{1}|0\rangle&=\frac{m^{2}_{D}f_{D}}{m_{c}+m_{q_{1}}},
\label{D}
\end{align}
Substituting the matrix elements Eqs. (\ref{D}) and (\ref{FFp}) into Eq. (\ref{CF}) and taking account of Lorentz structures of the correlation functions, one gets the phenomenological part of two correlation functions, we can obtain the phenomenological part of two correlation functions in terms of the form factors as
\begin{equation}
\begin{aligned}
\Pi_{1\mu}(p,q)&=-\frac{2f_{+}(q^{2})p_{\mu}+[f_{+}(q^{2})+f_{-}(q^{2})]q_{\mu}}{m_{D}^{2}-(p+q)^{2}}\frac{m^{2}_{D}f_{D}}{m_{c}+m_{q_{1}}}+\int_{s^h_{0}}^{\infty}\frac{\rho_\mu(s)ds}{s-(p+q)^{2}},\\
\Pi_{2\mu}(p,q)&=\frac{-[q^{2}2p_{\mu}-2q_{\mu}(p\cdot q)]}{m_{D}^{2}-(p+q)^{2}}\frac{if_{T}}{m_{D}+m_{P}}\frac{m^{2}_{D}f_{D}}{m_{c}+m_{q_{1}}}+\int_{s^h_{0}}^{\infty}\frac{\rho_\mu(s)ds}{s-(p+q)^{2}}.
\end{aligned}
\end{equation}
\par
Since calculation of the contributions from higher excited and continuum states on the phenomenological side is highly nontrivial, one can invoke quark-hadron duality to parameterize the corresponding spectral integral \cite{Belyaev:1993wp,Shifman:1980dk}:
\begin{align}
\int_{s_{0}^h}^{\infty}ds\frac{\rho_\mu(s)}{s-(p+q)^{2}}\simeq \int_{s_{0}}^{\infty}ds\frac{1}{\pi}\frac{{\rm{Im}}\Pi_{\mu}^{pert}(s)}{s-(p+q)^{2}},
\end{align}
Here, $s_0^h$ denotes the hadronic continuum threshold (a physical parameter representing the onset of higher resonances), while $s_0$ denotes the duality threshold (an effective parameter introduced by the quark-hadron duality assumption). And $\rho_\mu(s)$ is the spectral density of the higher excited and continuum states.

\subsection{The theoretical side\label{sec3B}}
In the region of large space-like momentum, the computation is based on the expansion of the T-product in the correlation functions near the light-cone. By contracting $c$ and $\bar{c}$ quark fields, we obtain
\begin{equation}
\begin{aligned}
\Pi_{1\mu}(p,q)&=i\int d^{4}xe^{iqx}\langle P(p)|\bar{q}_{2}(x)\gamma_{\mu}(1+\gamma_{5})S^{c}(x,0)i(1+\gamma_{5})q_{1}(0)|0\rangle,\\
\Pi_{2\mu}(p,q)&=i\int d^{4}xe^{iqx}\langle P(p)|T\bar{q}_{2}(x)\sigma_{\mu\nu}(1+\gamma_{5})q^{\nu}S^{c}(x,0)i(1-\gamma_{5})q_{1}(0)|0\rangle,
\end{aligned}
\label{cft}
\end{equation}
where $S^{c}(x,0)$ is the propagator of the charm ($c$) quark 
\begin{align}
S^{c}(x,0)&=-i\int\frac{d^{4}k}{(2\pi)^{4}}e^{-ik\cdot x}\frac{\slashed{k}+m_{c}}{m^{2}_{c}-k^{2}},
\end{align}
\par
In the method of the LCSR, the non-vanishing matrix elements, which is exactly the light-cone DAs, are defined in terms of twist functions. As a result, the DA matrix element obtained takes the form of
\begin{align}
\langle P(p)|\bar{q}_{2}(x)\gamma_{\mu}\gamma_{5}q_{1}(0)|0\rangle&=-ip_{\mu}f_{P}\int^{1}_{0}due^{iup\cdot x}\phi(u),
\end{align}
By employing the LCDAs and performing some straightforward calculations, two correlation functions on the theoretical side are obtained as
\begin{equation}
\begin{aligned}
\Pi_{1\mu}&=2f_{P}m_{c}p_{\mu}\int^{1}_{0}du\frac{\phi(u)}{m^{2}_{c}-(q+up)^{2}},\\
\Pi_{2\mu}&=-if_{P}[2p_{\mu}q^{2}-2q_{\mu}(p\cdot q)]\int^{1}_{0}du\frac{\phi(u)}{m^{2}_{c}-(q+up)^{2}}.
\end{aligned}
\end{equation}
\subsection{The form factors\label{sec3C}}
Now, we are in the position to use the LCSR method for the correlation function on two sides.  Let the coefficients of Lorentz structures in the correlation functions (\ref{CF}) on phenomenological side in Sect. \ref{sec3A} be equal to that of corresponding Lorentz structures in the correlation functions (\ref{cft}) the theoretical side in Sect. \ref{sec3B}, we get 
\begin{equation}
\begin{aligned}
-\frac{2f_{+}(q^{2})p_{\mu}+[f_{+}(q^{2})+f_{-}(q^{2})]q_{\mu}}{m_{D}^{2}-(p+q)^{2}}\frac{m^{2}_{D}f_{D}}{m_{c}+m_{q_{1}}}&=2f_{P}m_{c}\int^{1}_{\Delta}du\phi(u)\frac{p_{\mu}}{m^{2}_{c}-(q+up)^{2}},\\
\frac{-i[q^{2}2p_{\mu}-2q_{\mu}(p\cdot q)]}{m_{D}^{2}-(p+q)^{2}}\frac{f_{T}}{m_{D}+m_{P}}\frac{m^{2}_{D}f_{D}}{m_{c}+m_{q_{1}}}&=-i\int^{1}_{\Delta}duf_{P}[2p_{\mu}q^{2}-2q_{\mu}(p\cdot q)]\frac{\phi(u)}{m^{2}_{c}-(q+up)^{2}},
\label{cf}
\end{aligned}
\end{equation}
where $\Delta$ defines the integration lower limit, depending on the duality threshold $s_{0}$ of the $D$ meson, through
\begin{align}
\Delta=\frac{\sqrt{(s_{0}-m_{P}^{2}-q^{2})^{2}+4(m_{c}^{2}-q^{2})m_{P}^{2}}-(s_{0}-m_{P}^{2}-q^{2})}{2m_{P}^{2}}. \notag
\end{align}
\par
In order to suppress the contribution of the higher states and continuum, we shall perform Borel transformation $B_{M^2}$ \cite{Shifman:1978bx,Colangelo:2000dp} on both sides of Eq. (\ref{cf}). In the present case, this amounts to employing the following relation:
\begin{equation}
\begin{aligned}
B_{M^2}\frac{1}{m_{D}^2-(q+p)^2}&=\frac{1}{M^2}e^{-\frac{m_{D}^2}{M^2}},\notag\\
B_{M^2}\frac{1}{m_{c}^2-(q+up)^2}&=\frac{1}{uM^2}e^{-\frac{m_{c}^2+u(1-u)p^2-(1-u)q^2}{uM^2}}
\end{aligned}
\end{equation}
where $M^2$ is the Borel parameter. Upon Borel transformation on Eq. (\ref{cf}), one can get three relations, from which one can solve the following three FFs 
\begin{equation}
\begin{aligned}
f_{+}(q^2)&=\frac{m_{c}+m_{q_{1}}}{m_{D}^{2}f_{D}}f_{P}m_{c}\int^{1}_{\Delta}\frac{\phi(u)}{u}due^{FF},\\
f_{-}(q^2)&=-\frac{m_{c}+m_{q_{1}}}{m_{D}^{2}f_{D}}f_{P}m_{c}\int^{1}_{\Delta}\frac{\phi(u)}{u}due^{FF},\\
f_{T}(q^2)&=(m_{D}+m_{P})\frac{m_{c}+m_{q_{1}}}{m_{D}^{2}f_{D}}f_{P}\int^{1}_{\Delta}\frac{\phi(u)}{u}due^{FF},
\label{ff}
\end{aligned}
\end{equation}
with
\begin{align}
FF=&-\frac{1}{uM^{2}}(m^{2}_{c}+u\bar{u}p^{2}-\bar{u}q^{2})+\frac{m^{2}_{D}}{M^{2}}.
\end{align}
Simple comparison of the three relations (\ref{ff}) enable us to obtain two relations associated with three form factors $f_{+}^{P}$ , $f_{-}^{P}$ and $f_{T}^{P}(q^{2})$:
\begin{equation}
\begin{aligned}
f_{-}(q^2)&=-f_{+}(q^2),\notag\\
f_{T}(q^2)&=\frac{m_{D}+m_{P}}{m_{c}}f_{+}(q^2).
\end{aligned}
\end{equation}

\subsection{Branching fractions}
Equipped with the FFs derived above, we can now compute the differential decay widths for the processes under consideration. Specifically, for the semileptonic decay $D\rightarrow Pl\bar{v}_{l}$, which is the focus of this work, the LCSR framework allows us to express the differential decay width in terms of the FFs as follows:
\begin{equation}
\frac{d\Gamma(D\rightarrow Pl\bar{v}_{l})}{dq^{2}}=\frac{G_{F}^{2}|V_{cd}|^{2}}{24\pi^{2}}[(\frac{m_{D}^{2}+m_{P}^{2}-q^{2}}{2m_{D}})^{2}-m_{P}^{2}]^{\frac{3}{2}}|f_{+}(q^{2})|^{2},
\label{dbf}
\end{equation}
where $G_{F}$ is Fermi coupling constant, and $|V_{cd}|$ CKM matrix element. One sees that the differential decay width is proportional to the FFs squared. 
\section{Numerical analyses and discussion}
\subsection{Input parameters}
Before going to numerical analyses of the form factors and differential branching fractions, let us discuss briefly discuss the parameters involved in these semileptonic decay processes. For this purpose, we shall use the following inputs  \cite{ParticleDataGroup:2020ssz} for the parameters of the form factors and branching fractions , which are shown collectively in Table \ref{T3}.
\begin{table}[H]
	\centering
\caption{Parameters inputs for form factors and branching fractions \cite{ParticleDataGroup:2020ssz}}
	\begin{tabular}{p{2.5cm}<{\centering}p{2.5cm}<{\centering}p{2.5cm}<{\centering}p{3.6cm}<{\centering}p{3.6cm}<{\centering}}
        \hline
        \hline
        $m_{D^{0}}$ [GeV] &$m_{D^{+}}$[GeV] &$|V_{cd}|$&$ \tau_{D^{+}}$[$10^{-12}$s] &$\tau_{D^{0}}$[$10^{-13}$s] \\
		
        $1.865$ & $1.870$ &$0.220$ &$1.040\pm0.007$  & $4.101\pm0.015$ \\
        \hline
        
        $m_{u}$[MeV] & $m_{c}$[GeV] & $f_{\pi}$[MeV]  & $f_{D}$[MeV] & $f_{B}$[MeV]\\
        
        $2.16$ &$1.27$ &$130.3$ & $205.4$ &$190$\\
        \hline
        
        $G_{F}[\rm{GeV}^{-2}]$ &$ m_{e}$[GeV] &$ m_{\mu}$[GeV]  & $m_{d}$[MeV] & $m_{\pi^{\pm(0)}}$[MeV]\\
        
        $1.1664\times 10^{-5}$ &$0.511\times 10^{-3}$ &$0.106$  &$4.67$ & 139.57(134.98)\\
		\hline
        \hline
        \label{T3}
	\end{tabular}
\end{table}
The duality threshold $s_{0}$ has been explored and estimated in several effective scenarios \cite{Chernyak:1990ag,Dosch:2002rh,Matheus:2002nq,Bracco:2004rx,Lucha:2009uy}. Since $s_{0}$ is near to the squared mass of the lowest pseudoscalar $D$ meson, we  adopt $s_{0}=5.37\ \rm{GeV}^{2}$ corresponding to the $D$ channel. Despite the form factors should be independent of the Borel parameter ideally, the Borel mass must, in practice, be chosen within a window that satisfies the following criteria:
\begin{enumerate}
\item the contributions from the higher excited and continuum states are sufficiently  suppressed.
\item the form factors exhibit weak dependence on the Borel parameter.
\end{enumerate}
Given all these parameters in Table III, one can use Eq. (\ref{ff})  to perform calculation of the LCSR form factor $f_{+}(q^{2}=0)$ and obtain the Borel parameter ($M^2$) dependence of LCSR form factor $f_{+}(q^{2}=0)$. 
\par
\begin{figure}[H]
\centering
\includegraphics[height=6.5cm,width=8.5cm]{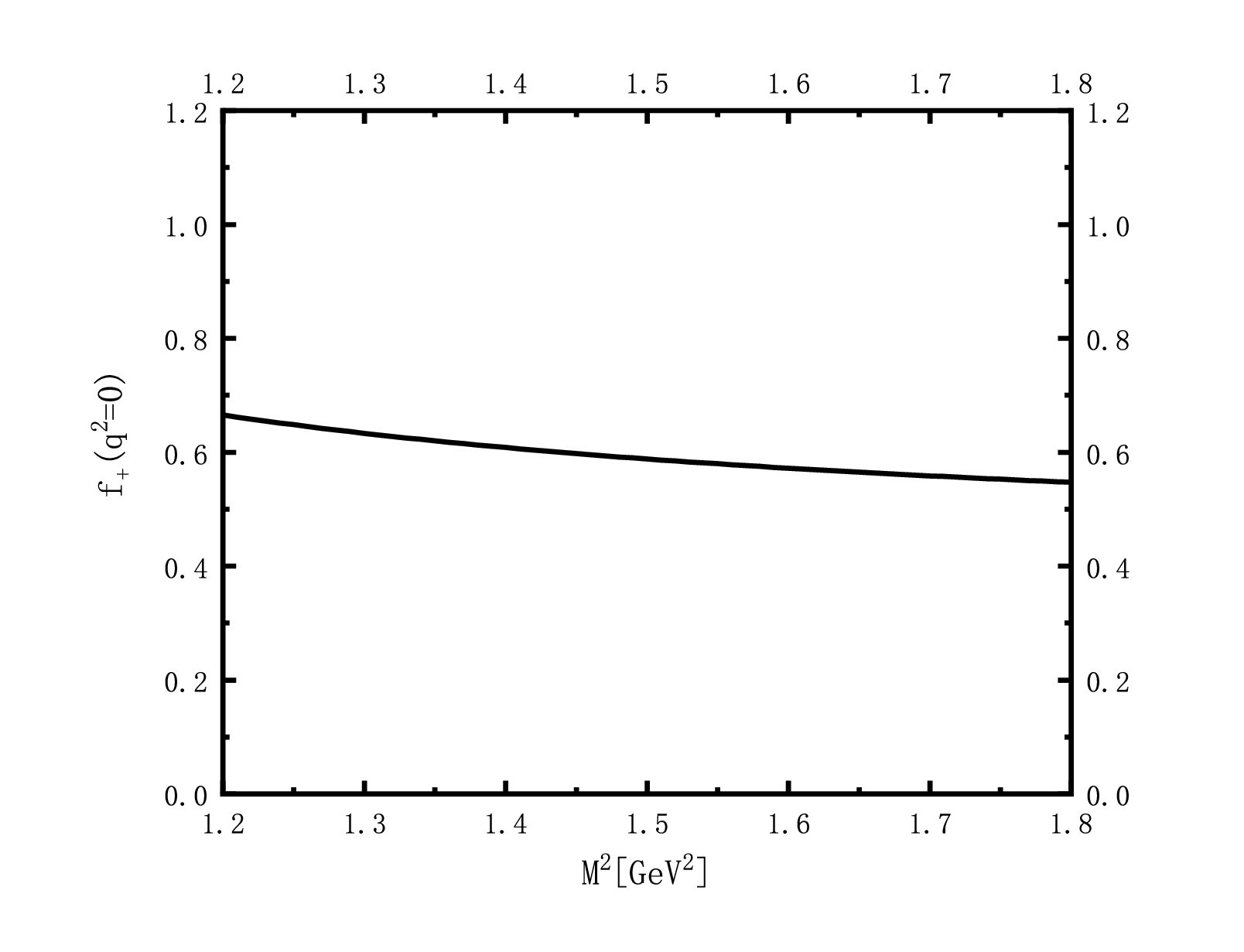}
\caption{$ D^{+(0)}\rightarrow \pi^{0(-)} $ transition form factor function on Borel parameter $M^{2}$ at threshold $s_{0}=5.37\ \rm{GeV}^{2}$ and $q^{2}=0$.}
\label{f-B}
\end{figure}
\begin{figure}[H]
\centering
\includegraphics[height=6.5cm,width=8.5cm]{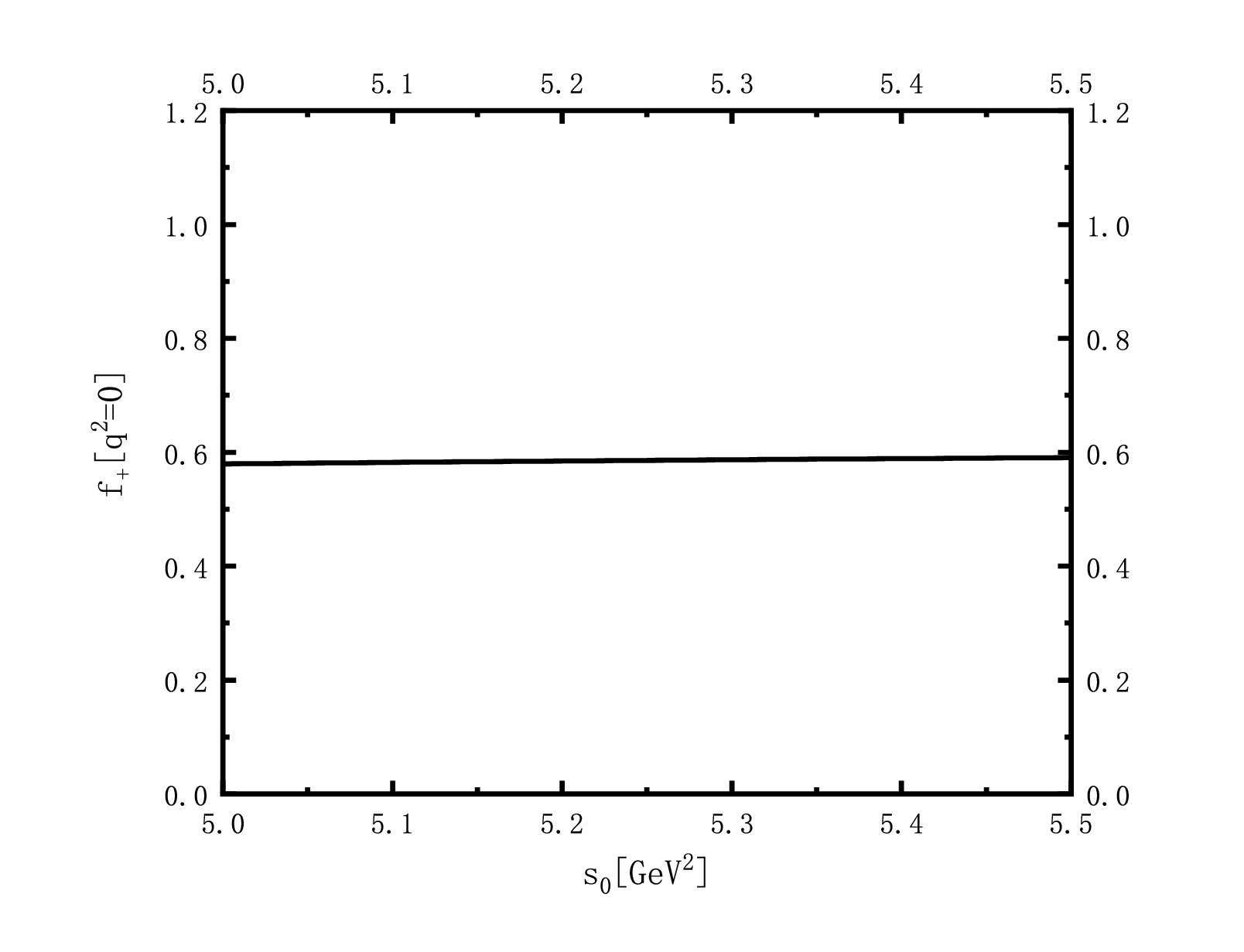}
\caption{$ D^{+(0)}\rightarrow \pi^{0(-)} $ transition form factor function on threshold $s_{0}$ at Borel parameter $M^{2}=1.5\ \rm{GeV}^{2}$ and $q^{2}=0$.}
\label{f-s}
\end{figure}

With these numerical analyses, one can numerically determine the thresholds and Borel windows for various semileptonic decay process in this work. In FIG.~\ref{f-B} and FIG.~\ref{f-s}, we plot the numerical predictions of the prediction of the form factors for $ D^{0}\rightarrow \pi^{-}$. As seen in FIG.~\ref{f-B} $f_{+}$  demonstrates a high stability against changes of $M^{2}$ with the range of $1.2\ \rm{GeV}^{2}<M^{2}<1.8\ \rm{GeV}^{2}$. This range is to be chosen as the optimized working interval. As shown in FIG. ~\ref{f-s}, the dependence of the form factors for $D^{+(0)}\rightarrow \pi^{0(-)}$ upon $s_{0}$ is quite weak and corresponding range of the  $s_{0}$ is to be chosen as the optimized working interval. 
Furthermore, we list, in Table \ref{T4}, our computed value of form factors and compare it with that obtained via other approaches for the decay $D\rightarrow \pi$ at $q^{2}=0$. For comparison, we have also computed the $B\to\pi$ form factors using the same BLFQ-based DA and LCSR framework. The results are listed in Table~IV along with other theoretical predictions.

\begin{table}
\begin{centering}
 \begin{threeparttable}
  \caption{Comparison of form factors in this work with that by other theoretical methods for $D/B\rightarrow P$. The first error is due to the Borel parameter $M^2$, and the second is due to the duality threshold $s_0$.}
  \begin{tabular}{p{6.5cm}<{\centering}p{2.4cm}<{\centering}p{2.4cm}<{\centering}p{2.4cm}<{\centering}p{2.4cm}<{\centering}}
\hline
\hline
Processes & \multicolumn{2}{c}{$B\rightarrow \pi$} &  \multicolumn{2}{c}{$D\rightarrow \pi$} \\
\hline
Form factor&$f_{+}$&$f_{-}$&$f_{+}$&$f_{-}$\\
\hline
This work&$0.27^{+0.005+0.004}_{-0.005-0.004}$&$-0.27^{+0.005+0.004}_{-0.005-0.004}$&$0.59^{+0.07+0.0057}_{-0.05-0.0106}$&$-0.59^{+0.07+0.0057}_{-0.05-0.0106}$\\

LCSR \cite{Khodjamirian:2000ds}&0.28&~&0.65&~\\

LCSR \cite{Khodjamirian:2011ub}&0.281&~&~&~\\

LCSR \cite{Ball:2006yd}&~&~&0.63&~\\

LCSR \cite{Li:2012gr}&0.28&~&0.62&~\\

LCSR \cite{Ball:2004ye}&0.258&~&~&~\\

LCSR \cite{Duplancic:2008zz}&0.26&~&~&~\\

LCSR \cite{Khodjamirian:2009ys}&~&~&0.67&~\\
\hline
LQCD \cite{Abada:2000ty}&~&~&0.57&~\\

LQCD \cite{FermilabLattice:2004ncd,Bernard:2009ke}&~&~&0.64&~\\

LQCD \cite{Al-Haydari:2009kal}&~&~&0.74&~\\

LQCD \cite{DiVita:2010mlb}&~&~&0.65&~\\

LQCD \cite{Na:2011mc}&~&~&0.666&~\\

LQCD \cite{FermilabLattice:2019ycs}&~&~&0.625&~\\
\hline
HQEFT$^{a}$ \cite{Wang:2002zba}&0.35&~&0.67&~\\

QCDSR$^{b}$ \cite{Ball:1993tp}&0.26&~&0.5&~\\
\hline
RQM$^{c}$ \cite{Faustov:2019mqr}&~&~&0.640&~\\

CCQM$^{d}$ \cite{Ivanov:2019nqd}&~&~&0.63&~\\

CLFQM$^{e}$ \cite{Cheng:2017pcq}&~&~&0.66&~\\
\hline
BES\uppercase\expandafter{\romannumeral3} \cite{BESIII:2017ylw}&~&~&0.622&~\\

BES\uppercase\expandafter{\romannumeral3} \cite{BESIII:2019qci,BESIII:2018xre}&~&~&0.635&~\\

HFLAV \cite{HFLAV:2019otj}&~&~&0.635&~\\

ETM \cite{Lubicz:2018rfs}&~&~&0.612&~\\
\hline
\hline
   \begin{tablenotes}
   \footnotesize
   \item[a] the Effective Field Theory of Heavy Quark
   \item[b] the QCD Sum Rule
   \item[c] the Relativistic Quark Model
   \item[d] the Covariant Confining Quark Model
   \item[e] the Covariant Light-Front quark Model
   \end{tablenotes}
  \end{tabular}
  \label{T4}
 \end{threeparttable}
\end{centering}
\end{table}

\subsection{The differential decay width}
\begin{figure}[H]
\centering
\includegraphics[height=6.5cm,width=8.5cm]{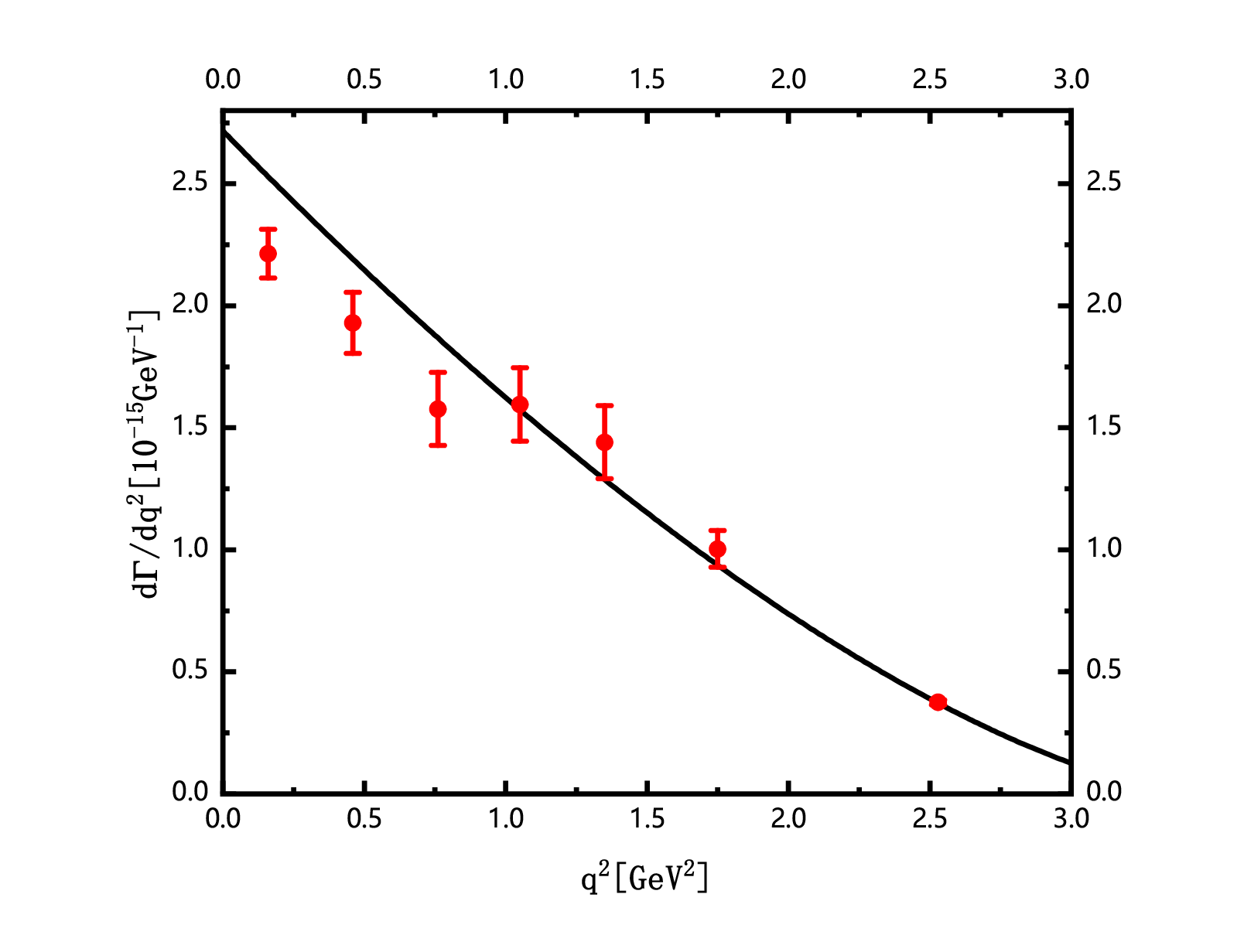}
\caption{The differential decay width for $D^{+}\rightarrow \pi^{0}e^{+}\nu_{e}$ as a function of $q^{2}$. The data (the dots) reported by BES\uppercase\expandafter{\romannumeral3} \cite{BESIII:2017ylw} are compared to prediction (solid line) by the obtained DA based on BLFQ. }
\label{G}
\end{figure}

Further, we employ the obtained FFs in Sect. \ref{sec3C} to compute the differential decay width by Eq. (\ref{dbf}). The numerical results for the differential decay width as a function of the transfer momentum $q^2$ is plotted in FIG.~\ref{G}.
In FIG.~\ref{G}, we present comparison of the differential decay width $d\Gamma/dq^2$ for $D^{+}\rightarrow \pi^{0}e^{+}\nu_{e}$ between our LCSR predictions (solid line) and the BES\uppercase\expandafter{\romannumeral3} data \cite{BESIII:2017ylw} (dots). Excellent agreement is observed in the high-$q^2$ region ( $q^{2}> 1\ \rm{GeV^2}$), where the LCSR framework is valid. However, a notable discrepancy emerges at $q^{2}< 1\ \rm{GeV^2}$, attributed to the the breakdown of the light-cone expansion in the low-momentum-transfer regime. We note that this discrepancy is consistent with theoretical expectations, as the LCSR formalism is built upon the assumption of large momentum transfer $(q^{2} \gg \Lambda^{2}_{\rm{QCD}}$, where $\Lambda^{2}_{\rm{QCD}}\sim 0.04-0.09 \ {\rm{GeV}^{2}})$. Such limitations are inherent to this theoretical approach and underscore the necessity of including higher-order corrections.
\par
The quoted uncertainties stem from variations in the Borel parameter $M^2$ and the duality threshold $s_0$, while the leading-twist distribution amplitude $\phi_{P}$ incorporates contributions from the $|q\bar{q}\rangle$ and $|q\bar{q}g\rangle$ Fock states.  To improve our prediction in this work, future refinements should systematically include additional effects, such as higher Fock states (e.g., $|q\bar{q}gg\rangle$), higher-twist light-cone distribution amplitudes, and next-to-leading-order corrections to the light-cone operator product expansion.

\section{Summary}
In this work, we have computed the transition form factors (FFs) for the semileptonic decays $D\rightarrow\pi$ using QCD light-cone sum rules (LCSR) with chiral currents, with focus on the role of pion light-cone distribution amplitudes (LCDAs). The twist-2 pion LCDA is systematically analyzed using both Gegenbauer polynomial expansions and the Basis Light-Front Quantization (BLFQ) method. The BLFQ-derived LCDA, which incorporates contributions from the $|q\bar{q}\rangle$ and $|q\bar{q}g\rangle$ Fock states, converges rapidly to the asymptotic form $\phi(u)=6u(1-u)$, yielding Gegenbauer coefficients of $a_{2}=-0.0283$ and $a_{4}=0.0130$ at $\mu=1\ \rm{GeV}$. Comparisons with previous theoretical predictions (e.g., QCD sum rules and lattice QCD) are given, underscoring the significance of nonperturbative dynamics of the strong interaction. 
\par
The $D\rightarrow\pi$ transition form factors $f_{+}$, $f_{-}$ and $f_{T}$,  derived from the LCSR framework, reveal two simplified relationships: $f_{-}=-f_{+}$ and $f_{T}=\frac{m_{D}+m_{\pi}}{m_{C}}f_{+}$, which are consistent with chiral symmetry. At $q^{2}=0$, our numerical result $f_{+}(0)=0.59$ agrees with experimental measurements from BES\uppercase\expandafter{\romannumeral3} (0.622-0.635) \cite{BESIII:2017ylw,BESIII:2019qci,BESIII:2018xre} and lattice QCD (0.57–0.74) \cite{Abada:2000ty,FermilabLattice:2004ncd,Bernard:2009ke,Al-Haydari:2009kal,DiVita:2010mlb,Na:2011mc,FermilabLattice:2019ycs}. The stability of the obtained FFs against the Borel parameter ($1.2\ \rm{GeV^2}<M^2<1.8\ \rm{GeV^2}$) and duality threshold ($s_{0}=5.37\ \rm{GeV^2}$) is verified via systematic uncertainty analyses.
\par
Our predictions for the differential decay width $d\Gamma/dq^2$ show good agreement with BES\uppercase\expandafter{\romannumeral3} data in the high-$q^2$ region ($> 1\ \rm{GeV^2} $), while deviations at low $q^2$ are attributed to the validity domain of LCSR $(q^{2} \gg \Lambda^{2}_{\rm{QCD}}$. The theoretical uncertainties may stem from neglected higher-twist corrections, omitted Fock states, and the truncation of the Gegenbauer expansion. These results offer promising prospects for extracting the CKM matrix element e.g., $|V_{cd}|$ ) via Eq. (\ref{dbf})  with high precision and exploring new physics beyond the Standard Model through precision tests of heavy-to-light decays.

\section*{ACKNOWLEDGEMENTS}
D. G acknowledges the support of the China Scholarship Council program (Project ID: 202510530001). D. J is supported by National Natural Science Foundation of China under Grant No. 12165017. Y. S is supported by Natural Science Foundation of China under Grant Nos. 11365018, 11375240 and 11565023. J. L is supported by Special Research Assistant Funding Project, Chinese Academy of Sciences, by the Natural Science Foundation of Gansu Province, China, Grant No.23JRRA631, and by National Natural Science Foundation of China, Grant No. 12305095. X. Z is supported by new faculty startup funding by the Institute of Modern Physics, Chinese Academy of Sciences, by Key Research Program of Frontier Sciences, Chinese Academy of Sciences, Grant No. ZDBS-LY-7020, by the Natural Science Foundation of Gansu Province, China, Grant No. 20JR10RA067, by the Foundation for Key Talents of Gansu Province, by the Central Funds Guiding the Local Science and Technology Development of Gansu Province and by the Strategic Priority Research Program of the Chinese Academy of Sciences, Grant No. XDB34000000. This research is supported by Gansu International Collaboration and Talents Recruitment Base of Particle Physics (2023-2027), and supported by the International Partnership Program of Chinese Academy of Sciences, Grant No.016GJHZ2022103FN. A portion of the computational resources were also provided by Gansu Computing Center and by Sugon Computing Center in Xi'an.

\end{document}